\documentclass[12pt,nofootinbib]{article}
\usepackage{amssymb}
\usepackage{graphicx}
\usepackage{graphics}
\usepackage{amsmath}
\usepackage{latexsym}
\def\const{\mbox{const}}

\def\l{\left(}
\def\r{\right)}

\newcommand{\be}{\begin{equation}}
\newcommand{\ee}{\end{equation}}
\newcommand{\ba}{\begin{align}}
\newcommand{\ea}{\end{align}}
\newcommand{\bg}{\begin{gather}}
\newcommand{\eg}{\end{gather}}
\newcommand{\bseq}{\begin{subequations}}
\newcommand{\eseq}{\end{subequations}}

\makeatletter
\def\gsim{\compoundrel>\over\sim}
\def\lsim{\compoundrel<\over\sim}
\def\compoundrel#1\over#2{\mathpalette\compoundreL{{#1}\over{#2}}}
\def\compoundreL#1#2{\compoundREL#1#2}
\def\compoundREL#1#2\over#3{\mathrel
         {\vcenter{\hbox{$\m@th\buildrel{#1#2}\over{#1#3}$}}}}
\makeatother

\begin{document}

\title{SUSY in the sky \\ or \\ keV signature of sub-GeV gravitino dark matter}

\author{S.~V.~Demidov$^{1,}$\thanks{{\bf e-mail}: demidov@ms2.inr.ac.ru}\,,
D.~S.~Gorbunov$^{1,2,}$\thanks{{\bf e-mail}: gorby@ms2.inr.ac.ru}\,,
\\
$^{1}${\small{\em
Institute for Nuclear Research of the Russian Academy of Sciences,
}}\\
{\small{\em
60th October Anniversary prospect 7a, Moscow 117312, Russia
}}\\
$^{2}${\small{\em
Moscow Institute of Physics and Technology,
}}\\
{\small{\em
Institutsky per. 9, 
  Dolgoprudny 141700, Russia
}}\\
}
\date{}

\maketitle

\vspace{-10cm}
\begin{flushright}
INR-TH-2014-010
\end{flushright}
\vspace{8.5cm}

\begin{abstract}
We point out that recently discovered 3.5~keV line in X-ray spectra
from various galaxy clusters and the Andromeda galaxy can be naturally
explained by physics of a sector responsible for spontaneous
supersymmetry breaking in models with R-parity. In this scenario the
source of this line could be decay of sgoldstino - scalar superpartner
of massive gravitino. At the same time the dominant dark matter
component is stable gravitino whose mass is predicted to be about
0.25~GeV. 
\end{abstract}

\section{Introduction}

Analysis of X-rays coming from galaxy clusters and from the Andromeda
galaxy has revealed~\cite{Bulbul:2014sua,Boyarsky:2014jta} an
anomalous signal at energy $E\approx 3.5$~keV. Although the
observation has significance only at $(3\sigma-4\sigma)$ level and its
explanation in the framework of known physics is possible it would be
very exciting if the source of this line is in new physics probably
related to so far elusive dark matter. Various scenarios have been
studied including sterile
neutrinos~\cite{Ishida:2014dlp,Abazajian:2014gza,Baek:2014qwa}, XDM
model~\cite{Finkbeiner:2014sja,Frandsen:2014lfa}, axion-like
particles~\cite{Higaki:2014zua,Jaeckel:2014qea,Lee:2014xua,Nakayama:2014ova,Nakayama:2014cza},
millicharged particles~\cite{Aisati:2014nda} and light
inflaton~\cite{Bezrukov:2014nza}. Below we are interested in possible
explanations of this phenomena in the framework of supersymmetric
theories. On this way there were considered scenarios with R-parity
violation, where very long-lived dark matter particle of 7\,keV decays
into neutrino and photon, thus explaining the line at 3.5\,keV. Light
gravitino explanation~\cite{Bomark:2014yja} (see
also~\cite{Kolda:2014ppa}) needs quite low reheating
temperature. Models with light
axinos~\cite{Kong:2014gea,Choi:2014tva,Liew:2014gia} are also quite
tricky because they require quite small (less than 10~GeV) bino mass
and other neutralinos to be considerably heavier.

In this paper we seek for the explanation in supersymmetric models
{\em with R-parity}, so both proton and lightest supersymmetric
partner are stable. We propose that sector responsible for spontaneous
supersymmetry breaking (SSB) accounts for the 3.5~keV line and at the
same time for dark matter. This sector contains goldstino, which
becomes longitudinal component of gravitino in supergravity, and its
superpartner(s). The latter in the simplest case are spin-0 bosons and
are called sgoldstinos (for brief review on phenomenology of goldstino
supermultiplet see \cite{Gorbunov:2001pd}). We assume that the
pseudoscalar sgoldstino (axion-like particle) has mass around 7~keV and gives
observed in X-rays signal due to its decays into pair of photons
(sgoldstinos are R-even and hence unstable). We
calculate thermal production of sgoldstinos in early Universe and find
that if we require that their two-photon decays explains 3.5~keV line
in X-ray spectra sgoldstinos are always subdominant part of the dark
matter. At the same time the dominant component is gravitino and it is
interesting that the gravitino mass is also fixed in this
setup and is about 0.25~GeV.


\section{Setup}

We start with presenting relevant interaction lagrangian (see
e.g.~\cite{Perazzi:2000id,Gorbunov:2000th}) of pseudoscalar
sgoldstino with gluons and photons
\be
\label{lagrangian}
{\cal L}_P = \frac{M_{3}}{2\sqrt{2}F} PG^{a}_{\mu\nu}
\tilde{G}^{a\mu\nu} + \frac{M_{\gamma\gamma}}{2\sqrt{2}F}
PF_{\mu\nu}\tilde{F}^{\mu\nu}\,, 
\ee
where $\tilde{F}_{a}^{\alpha\mu\nu} =
\frac{1}{2}\epsilon^{\mu\nu\lambda\rho} F^{\alpha}_{a\lambda\rho}$ and 
$M_{\gamma\gamma} = M_{1}\cos^2{\alpha_W} + M_{2}\sin^2{\alpha_W}$, 
$M_{1,2,3}$ are soft gaugino masses and $\sqrt{F}$ is supersymmetry
breaking scale. Sgoldstinos also interact with 
quarks, leptons and other gauge fields, that we address later.  
Gravitino couples to other fields mostly through the longitudinal
component, goldstino. Its interactions  
with other fields are given by derivative coupling to supercurrent
\be
\label{lagrangian-2}
{\cal L}_{3/2} = \frac{1}{F}\partial_{\mu}\psi_{3/2}J^{\mu}_{\rm SUSY}\,.
\ee
This approximation is valid up to energies
$E\sim\sqrt{F}$. Sgoldstino with mass around $7$~keV decays mainly
into pair of photons and its lifetime is
\be
\label{lifetime}
\tau_P = \frac{32\pi F^2}{M^2_{\gamma\gamma}m_P^3} \approx 
1.9\times 10^{27}~{\rm s} 
\l \frac{\rm 1~TeV}{M_{\gamma\gamma}}\r^2
\l \frac{\rm 7~keV}{m_P}\r^3
\l \frac{\sqrt{F}}{\rm 10^{10}~GeV} \r^4
\ee
If sgoldstino composes only a part of dark matter contributing a 
fraction $\epsilon$ to the present dark matter mass density, then 
the requirement that sgoldstino two-photon decay 
explains 3.5~keV X-ray line yields the following
constraint  
\be
\label{constraint}
\tau_P \sim \epsilon\times\left(4\times 10^{27} - 4\times
10^{28}\right)~{\rm s}.
\ee
The fraction $\epsilon$ should be not much smaller than $10^{-10}$,
otherwise $\tau_S$ is less than the age of the Universe and
all relic sgoldstinos decay by present time. This implies the lower bound on
supersymmetry breaking scale 
\be
\sqrt{F}
\gsim 3\times 10^7~{\rm GeV}.
\ee


\section{Relic abundances of sgoldstino and gravitino}

Let us continue with description of sgoldstino production in the early 
Universe. Assuming that the temperature is below the supersymmetry
breaking scale $\sqrt{F}$ we can expect that the main sources of this
production are $2\to 2$ reactions involving gluons $g$, gluinos
$\tilde g$, quarks $q$ and squarks $\tilde q$ such as $gg\to gP$,
$qg\to qP$, $\tilde{q}g\to\tilde{g}P$ etc. Boltzmann equation which
describes evolution of sgoldstino number density reads 
\be
\label{boltzman}
\frac{dn_P}{dt} + 3Hn_P = \gamma_P,
\ee
where $H$ is Universe expansion rate and sgoldstino 
production rate $\gamma_P$ has a form similar to that of the
axion thermal production (see e.g. Ref.~\cite{Graf:2010tv})
\be
\gamma_P \approx \frac{\alpha_sM_3^2T^6\zeta(3)}{\pi^2F^2}\beta,
\ee
where $\beta$ is a factor of order unity and which has negligibly
small dependence on plasma temperature $T$. For the axion-like
particle production in the Standard Model this factor was estimated in
the Hard Thermal Loop approximation~\cite{Graf:2010tv} and more
recently and accurately in~\cite{Salvio:2013iaa} extending previous
results to larger values of gauge coupling constants. In what follows we take
$\beta\approx 1$, so that numerical estimates of model parameters
below are accurate up to a factor of few, which seems reasonable given
the uncertainty in lifetime \eqref{constraint} following from 
observations~\cite{Bulbul:2014sua,Boyarsky:2014jta}. 
Using conservation of the entropy in co-moving volume, $s\,a^3=\const$ and
relation $H=T^2/M^{*}_{Pl}$, where $M_{Pl}^{*} \approx
M_{Pl}/1.66\sqrt{g_*}$ and $g_*\approx 229$ is effective number of
degrees of freedom in supersymmetric plasma at high temperatures, 
Eq.~\eqref{boltzman} can be cast into  the form
\be
\label{B-eq}
\frac{d}{dT}\left(\frac{n_P}{s}\right)
= -\frac{\gamma_PM_{Pl}^*}{sT^3}.
\ee
The r.h.s of this equation does not depend on the temperature apart
from mild changes in $g_*$ at particle thresholds and running of 
strong coupling constant entering $\gamma_P$. Hence solution of
Eq.\,\eqref{B-eq} can be written as 
\be
\frac{n_P(T)}{s(T)} =
\l\frac{\gamma_PM_{Pl}^*}{sT^3}\r_{T=T_R}\cdot(T_R - T)
\ee
where $T_R$ is the reheating temperature. At sufficiently low
temperatures obeying $m_{soft}\ll T \ll T_R$ where $m_{soft}$ is the
mass scale of superpartners one approximates 
\be
\frac{n_P(T)}{s(T)} =
\l\frac{\gamma_PM_{Pl}^*}{sT^3}\r_{T=T_R}\cdot T_R.
\ee
Then sgoldstino number density at present time is 
\be
n_{P,0} = \frac{s_0}{s(T)}\,n_{P}(T) =
n_{\gamma,0}\,\frac{g_*(T_0)}{g_*(T_R)}
\frac{\beta\alpha_s M_{Pl}^*M_{3}^2}{2F^2}\,T_R
\ee
and for sgoldstino relative contribution to the present energy density
of the Universe $\Omega_P$ one obtains following estimate 
\be
\label{sgold}
\Omega_Ph^2 \approx 1.7\times 10^{-8}\times
\l \frac{m_P}{7~{\rm keV}}\r
\l \frac{M_3}{3~{\rm TeV}}\r^2
\l \frac{T_R}{1.7\times 10^8~{\rm GeV}}\r
\l \frac{10^{10}~{\rm GeV}}{\sqrt{F}}\r^4,
\ee
where $h\simeq 0.7$ refers to the normalized Hubble parameter. 
Similar contribution of relic 
gravitinos $\Omega_{3/2}$ was estimated in~\cite{Bolz:2000fu},  
\be
\label{grav}
\Omega_{3/2}h^2 \approx 0.13\times\l \frac{M_3}{3~{\rm TeV}}\r^2
\l \frac{T_R}{1.7\times 10^{8}~ {\rm GeV}}\r \l \frac{10^{10}~ {\rm
    GeV}}{\sqrt{F}}\r^2 
. 
\ee

Comparing Eqs.~\eqref{sgold} and~\eqref{grav} one can see that
production of gravitinos is more effective and when mass density of
gravitinos is sufficient to explain the dark matter the abundance of
sgoldstino is quite small: they form a subdominant fraction of the
dark matter. Fraction $\epsilon\equiv
\Omega_P/\Omega_{3/2}$ of produced sgoldstino checks with what 
constraint~\eqref{constraint} gives if   
\be
\label{scale}
\sqrt{F} \sim (0.8-1.2)\times 10^9~{\rm GeV}\times
\l\frac{m_P}{7~{\rm keV}}\r^{2/3}
\l \frac{M_{\gamma\gamma}}{1~{\rm TeV}}\r^{1/3},
\ee
where we used the expression for sgoldstino lifetime~\eqref{lifetime}.
This implies gravitino mass $m_{3/2} =
\sqrt{\frac{8\pi}{3}}\frac{F}{M_{Pl}}\sim (0.15-0.35)$~GeV. At the
same time the requirement~\eqref{grav} that gravitino composes the
dark matter determines the value of the reheating temperature 
\[
T_R\sim \l 1.0-2.4\r\times 10^6~\text{GeV}\,.
\]
It exceeds the required value of the supersymmetry breaking
scale\,\eqref{scale}, which justifies the use of effective lagrangians
\eqref{lagrangian}, \eqref{lagrangian-2} 
to describe sgoldstino and gravitino production in the early Universe.


\section{Discussion and conclusions}
To summarise, we present a supersymmetric model with two-component dark
matter. The dominant part is gravitino with mass in sub-GeV region 
while the subdominant part is long-lived pseudoscalar sgoldstino. The
latter decays into two-photons and can give rise to the unexplained
excess in X-ray spectra at
3.5~keV~\cite{Bulbul:2014sua,Boyarsky:2014jta}. 
Note that we consider here
pseudoscalar sgoldstino only as an example and scalar sgoldstino of
3.5~keV is not a worse candidate.

Actually both sgoldstinos are produced in the early Universe at the
same rate provided they are lighter than $\sqrt{F}$. If both are
sufficiently long lived their present abundances (c.f.~\eqref{sgold})
are related as $\Omega_S = \frac{m_P}{m_S}\; \Omega_P$. If, say,
pseudoscalar sgoldstino explains observed X-ray signal and $m_S \gsim 
1$~MeV the scalar sgoldstino decays by the present epoch that may
interfere with primordial nucleosynthesys and/or distort CMB
spectrum. Lighter scalar sgoldstino contributes to present dark matter
and similar to pseudoscalar sgoldstino its decay into pair of photons
gives another  X-ray line at energy $E_{\gamma}=m_S/2$. Given the same
amount of sgoldstinos the expected X-ray fluxes from scalar
$\Phi_{\gamma}^{S}$ and pseudoscalar $\Phi_{\gamma}^{P}$ are related
as $\Phi_{\gamma}^{S}/\Phi_{\gamma}^{P} =  m_S^3/m_P^3$ (see
Eq.~\eqref{lifetime}). Observation of this second line provides an
additional support to the suggested explanation of the X-ray excess at
7~keV.

Light gravitino and sgoldstinos are natural in e.g. models with gauge
mediation of supersymmetry breaking
\cite{Giudice:1998bp,Dubovsky:1999xc} 
which elegantly circumvent experimental bounds on SM superpartner mass scale
from studies of flavor and CP-violation.  
However, the adopted hierarchy between sgoldstino and gravitino masses is
not typical. Let us illustrate this issue with a toy model of goldstino
superfield $X=\phi_X+\sqrt{2}\theta\psi_{3/2}+\theta^2F_X$ and   
superpotential $W=-FX$, which triggers supersymmetry breaking. The 
K\"aler potential is 
\be
{\cal K} = X^{\dagger}X - \frac{\alpha}{4M^2}\left(X^\dagger X\right)^2
- \frac{\beta}{6M^2}\left(X^\dagger X^3 + X^{\dagger3}X\right),
\ee
where the last two terms model corrections from hidden sector
and from mediator and gravity interactions. These terms are suppressed by some
energy scale $M$ and $\alpha,\beta$ are some constants.
After supersymmetry breaking $\langle F_X\rangle = F$ and decomposing
$\phi_X = \frac{1}{\sqrt{2}}\left(S + iP\right)$ one obtains for
sgoldstino masses
\be
m_S^2 = \frac{F^2}{M^2}\left(\alpha +\beta\right),\;\;\;\;
m_P^2 = \frac{F^2}{M^2}\left(\alpha-\beta\right).
\ee
Comparing them with gravitino mass one can see that given $M\lsim
M_{Pl}$ the hierarchy $m_P\ll m_{3/2}$ implies either a 
smallness of $\alpha$ and $\beta$ or strong degeneracy between their
values. Models of this kind have been considered for example in 
Refs.~\cite{AlvarezGaume:2010rt,AlvarezGaume:2011xv,AlvarezGaume:2011db} 
however in a different context. 
Dark matter gravitino of sub-GeV mass has been also discussed in literature,
see e.g.\,\cite{Feng:2008zza}. 
The values of model parameters required
for the suggested explanation of 3.5\,keV line are consistent with
present astrophysical, cosmological and phenomenological bounds on 
goldstino sector\,\cite{Gorbunov:2001pd}.  We have checked also that
for the scale of SUSY breaking~\eqref{scale} lifetime of 
superpartners of SM particles is small enough ($\lsim 0.1$~s) 
for these species to interfere with the primordial
nucleosyntesis. Successful production of gravitino and sgoldstino by
scattering of particles in primordial plasma implies the Universe
reheating temperature $T\sim 10^6$\,GeV, that seems to leave enough
room for baryogenesis in particular realizations of the suggested
scheme.   

\vskip 0.2cm 

We thank Fedor Bezrukov for useful comments and Zhaofeng Kang for
encouraging communication. 
The work is supported in part by the grant of the President of the
Russian Federation NS-2835.2014.2 and  by Russian Foundation for Basic
Research grants 13-02-01127a. The work of D.G. is partly 
supported by RFBR grant 14-02-00894a.


\end{document}